\documentclass[journal]{IEEEtran}

\usepackage[english]{babel}
\usepackage[utf8]{inputenc}
\usepackage{amsmath, subfigure, graphicx}
\usepackage[colorinlistoftodos]{todonotes}
\usepackage{amsmath, amssymb}
\usepackage{tikz}
\usepackage{float}
\usepackage{comment}
\usepackage{url}
\usepackage{xspace}
\usepackage{wrapfig}

\floatstyle{ruled}
\newfloat{program}{thp}{lop}
\floatname{program}{Program}

\title{
An Energy Management System Approach for Power System Cyber-Physical Resilience}

\author{
     Katherine R. Davis,~\IEEEmembership{Senior Member,~IEEE}

    \thanks {    
    K. R. Davis is with Texas A\&M University: katedavis@tamu.edu. Submitted to Virtual Workshop on Cyber Experimentation and Science of Security (CESOS) 2021}

}

\begin{document}
\maketitle

\begin{abstract}
Power systems are large scale cyber-physical critical infrastructure that form the basis of modern society. The reliability and resilience of the grid
%of grid physical infrastructure 
is dependent on the correct functioning of related subsystems, including computing, communications, and control.
%, which currently are considered separately despite the strong interdependencies.
%
The integration 
%of cyber communications and control systems into the power grid infrastructure 
is widespread and has a profound impact on the operation, reliability, and efficiency of the grid.  
%Cyber 
Technologies comprising these 
%efficient power system management 
infrastructure can 
%cause and/or 
expose new sources of threats.
Mapping these threats to their grid resilience impacts to stop them early requires a timely and detailed %understanding 
view of the entire cyber-physical system.
%needs to be identified and defended.  
Grid resilience must therefore be seen and addressed as a cyber-physical systems problem. 
%
%One important possible consequence is the introduction of cyber-induced or cyber-enabled disruptions of physical components. 
%
This short position paper presents several key preliminaries, supported with evidence from experience,
%of over a decade and several from our experience) 
to enable 
%these capabilities
%
cyber-physical 
situational awareness and 
intrusion response 
%for power grid %power system 
%resilience 
through a cyber-physical energy management system.   
\end{abstract}

%\begin{IEEEkeywords}
%Cyber-physical systems, cyber security, contingency analysis, operational reliability, cyber-physical topology
%\end{IEEEkeywords}

%%%% INTRODUCTION %%%%%%
\section{Introduction}
Power system infrastructures have long been rigorously studied and analyzed. Historically, however, their cyber systems, the interplay between cyber and physical, and the potential impacts of loss of trustworthiness at any point, have not been given the same level of attention.
The current need is that cyber-physical power systems must be able to ``ride through" extreme events, including 
%different types of 
human-perpetrated events, 
%including cyber-attack 
while maintaining their critical societal functions.
Part of the challenge is these events fall outside the scope of normal daily operations, so they require a new approach with new models and tools to address.
%-- we need to bridge the gap. 

Cyber-physical Resilient Energy Systems (CYPRES) \cite{cypresweb} is funded by the U.S. Department of Energy's Cybersecurity for Energy Delivery Systems (CEDS) to achieve 
%the objective in 
its name.  CYPRES aims to transform the energy sector's resilience capabilities in cyber-physical power systems by designing a new approach to energy management.
%system %approach 
%proof-of-concept that is truly cyber-physical and inherently considers security.
%
%A key function we focus on here is 
%Key functions driving its design include to 
A primary function is to \textbf{\textit{help electric utility industry stakeholders more effectively prevent and detect cyber-attacks that threaten their operational environment, and recommend controls, by coupling cyber-physical models and cyber-physical data.}}

CYPRES, as a new type of energy management system (EMS), would provide cyber-physical grid analyses for stakeholders over the entire life cycle of an event. 
%CYPRES supports all phases of the lifecycle of an event. %[figure] 
%What is a resilience life cycle? Resilience life cycle… 
The new EMS must include the modeling and data analytic capabilities 
%that enable stakeholders 
%to plan for and achieve resilience %throughout the entire \textbf{\textit{life-cycle of an event}}, 
%by informing how 
to prepare for, endure, and respond to events (threats), as well closing the loop to learn from these events (threats) and plan the system better.  
This resilience life cycle approach is critical to grid cyber-physical security.  
In fact, the importance of security for critical infrastructure is   
%(arguably) 
completely rooted in operational resilience.  This perspective is a key strategy: it is \textit{the} way to 
%develop and deploy 
create solutions that will be tractable: by prioritizing security based on resilience.

The CYPRES techniques are being developed and validated with respect to the following
threat model:
%, which from a resilience standpoint is critical as it includes detailed representation of the worst case scenarios: 
\textbf{\textit{Advanced cyber adversaries are targeting physical power system impact through multi-stage attacks.}} Such threats require cohesive use of cyber and physical data and models together to defend.
%
%The most important goal is that the system continues to operate reliably, which requires cyber and physical coordination: cyber actions must be power aware, and power actions must be cyber aware.
%
The elements of CYPRES help form the answer for how to 
%are necessary to be able to answer the question of how to use both cyber and physical information together for defense 
achieve this defense in the most effective coordinated manner.

\section{Cyber-Physical Resilience Framework}
\begin{figure*}
\centering
\includegraphics[scale=.5]{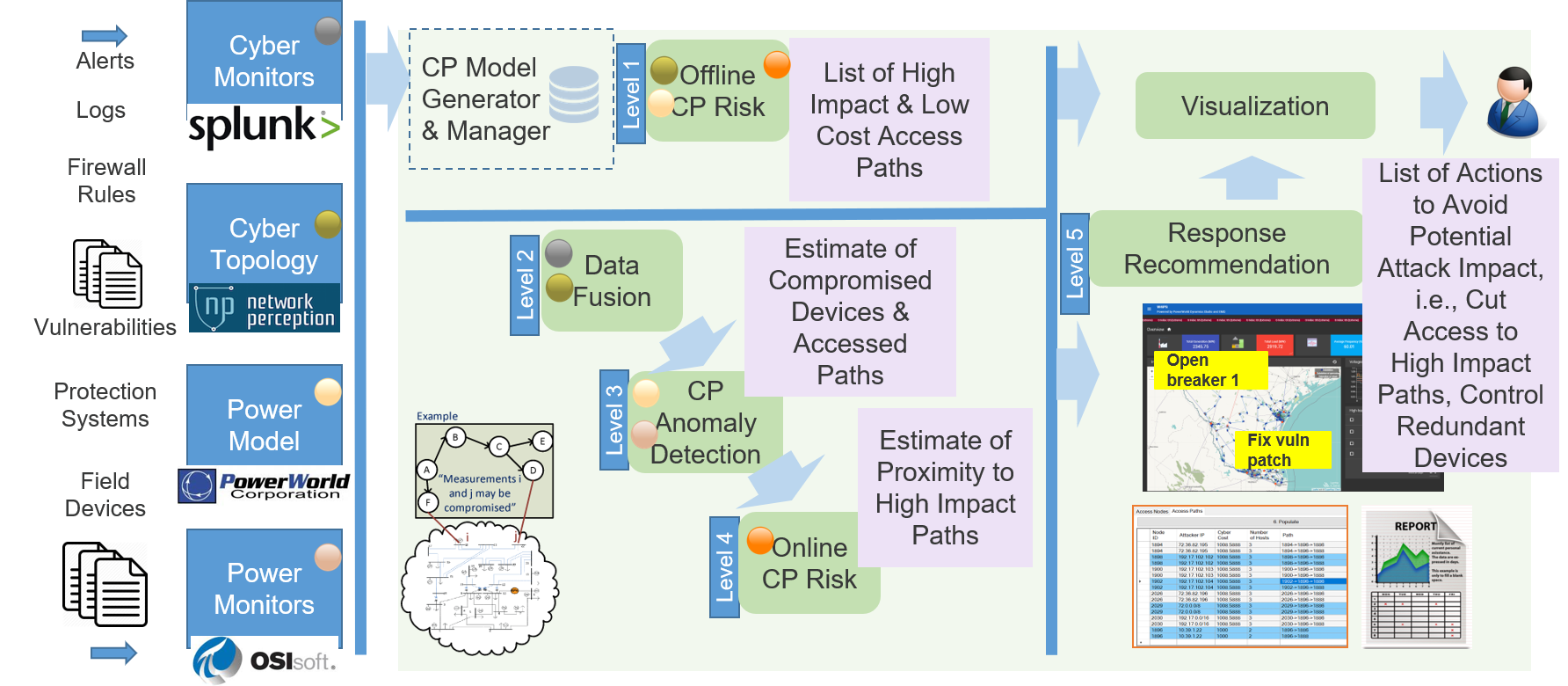}
\vspace{-0.2in}
\caption{The CYPRES model and example of the cyber-physical energy management system workflow.}
\label{fig:cypresworkflow}
% kate 9/13/14 the below was making the figure caption overwrite other text..
%\vspace{-0.2in}
\end{figure*}

CYPRES has been developing modeling and analytic techniques and tools that comprise 
%the elements of what is needed 
a proposed next generation energy management system. 
The framework is cyber-physical and inherently considers security and risk throughout the data flow pipeline (monitoring, to analysis, to control).
%
%[]details
%
The feasibility 
%of the CYPRES approach 
is supported by use of well-known (industry used) data sources, coupled with the analytics that form its functional modules (Fig.~\ref{fig:cypresworkflow}).

%CYPRES’s solution to cyber-physical 
%The solution to 
Resilience begins in the planning phase.
%
%The goal is improving cyber-physical situational awareness, which includes by offering preventative risk analysis of what is possible. 
The goal is always to catch and prevent potential events early, while intentionally planning for the inevitable imperfection of models and for the inevitable imperfection of security defenses.
%, but since that is not possible

%\begin{figure}
%\end{figure}
\noindent \textbf{Model fusion}. Hence, first
%the first part of the approach
%CYPRES approach, closing the loop with a unified model, 
is \textit{cyber-physical model fusion}, that combines cyber and physical model information.
To provide cyber-physical situational awareness, a map is needed, to help stakeholders understand the system.
The map includes the power system topology, the cyber network topology, and the part in between, where they intersect~\cite{weaver2016cyber, synthetic_comm}:
%.  In modeling terms, this requires the following. 
A \textit{cyber topology} of the control network protects and thus affects grid operational reliability; it 
%is  geographically distributed and 
encompasses control center and substation networks. 
%Cyber topology modeling is discussed in Section~\ref{sec:topology}.
%
A \textit{power topology} is how cyber-physical interactions 
%with signals and sensors 
map to physical impact;
%.    %Cyber-physical modeling is also discussed in Section~\ref{sec:topology}.
%The power 
topology and state are essential to quantifying impacts.
%in the threat model.
%such as line outages induced through cyber-attacks
A \textit{threat model} represents the characteristics of 
%the relevent 
cyber-physical threats that are critical to design and test of the defense tools. 
% Cite:
% Weaver 2016
% Wlazlo 2019
%
%
%
Part of the challenge is that these requirements are not historically well-known or documented. %However, going forward, 
This cannot be the case going forward, because a stakeholder needs to understand the system to best defend it. 
This major challenge is also where CYPRES helps: in identifying, mapping, and characterizing the system.
%se interconnections. 
%It is critical to shed light on and fortify the currently unseen cyber backbone and its interconnections to grid operational components.  The new
%CYPRES 
%EMS would directly provide this.

\noindent \textbf{Fused model mathematical representation}. The model generation step incorporates the topologies and threat information to create a state space representation for subsequent analyses. It is 
%created at the beginning and again 
updated when needed based on changes.
%to the system or threat information. 
The state space embeds physical impact/operational reliability metrics specific to the use case and threat model.
%s considered.

\noindent \textbf{Dataflow integrity and security monitoring}.
Detection is an indispensable component of situational awareness. Monitoring for resilience is based on dataflows and their functions.  %framework\name 
This flow-based approach directly supports the collection and correlation of data from cyber and physical components throughout the networks.
%, e.g., intrusion detection systems (IDSs) specifically designed to find malicious activities in power grid infrastructures. 
%In an online mode, 
This enables use of the generated models with available alerts to estimate the security state of the system.

\noindent \textbf{Data fusion for ML-enhanced CPS state estimation}. %
Next is fusion of cyber and physical network data through techniques with feature extraction 
%in support of 
enabling \textit{effective} 
%and \textit{accurate} %subsequent 
machine learning (ML)-based cyber-physical intrusion detection, inferencing, and/or state estimation. Alerts from sensors on the cyber side are correlated with data from the physical side to infer what is going on in the system~\cite{wlazlo_mitm, Fusion2021, a3d,boyaci2021graph}.

\noindent \textbf{Preventative risk analyses for ahead-of-time mitigations}. Fused models and data enables cyber-physical risk analyses~\cite{SOCCA2014,davis2015cyber,AmaraCPBC2021}, to find potential events %potentially induced by a cyber adversary 
and identify components that need more defense.
%security strengthening
%that are otherwise missed. 
Prevention also involves planning: learning from past events, current state, and fused models enables such studies. Specialized testbeds (such as our Resilient Energy Systems Lab \textit{RESLab}~\cite{Sahu2020}, a realistic emulated cyber-physical grid environment) support the realism of such studies without exposing or requiring the real stakeholder network. %stakeholders 

\noindent \textbf{Adaptive risk analyses for online decision support}.
No security is perfect. Regardless of how well a system is defended, response strategies are critical.
%and mitigation is critical. 
Such a response plan 
%with this resilience framework, can be 
is well-informed by the other components and can be designed to provide recommendations for response and mitigation.
%and intentionally designed.
%strategy is critical. Here we can take advantage of all of the offline capabilities described above, while also including an online analysis of the current state of the system. Then, we can see how the current state (what we actually see) matches up to high-risk cyber-physical threats identified in the planning phase (what the models predict that we should see).
%
%These analyses provide recommendations for response and mitigation.

%\section{Conclusion}
The above components are primary functions of resilience.
%:
%We now outline the functional block of resilience. 
%They can be divided into generation of the analysis model and core analysis based on the model. 
%There are several primary functions: 
%model generation, power systems analysis, cyber security state estimation, offline risk analysis, fusion-based detection and inference, online risk analysis, and coordinated cyber-physical response.  
%
%During operation, 
These would run periodically in the cyber-physical energy management system, updating as system state changes.
%of the networks change. 
%
%Ultimately, the core solution for grid resilience fuses information from the threat model, the cyber topology, and the power topology, and real time data to assess the system and provide early threat detection, and based on that assessment, it determines how best to remediate the situation with available cyber and physical controls to preserve grid operational functions.
%
%For these functions, 
With well-defined and consistent interfaces, each may be researched %, developed, 
and improved independently. Ideally, 
%It is %even 
%better if 
multiple (or all) functions 
%(or ideally, all functions) can 
should be 
%designed
%unified 
%and 
optimized together, e.g., reducing 
%additional 
computational burden %associated 
of modeling, monitoring, and control, and to simplify system defenses (including its design) based on key functions and their %cyber-physical 
interdependencies.
%associated with the data flows of those functions).  
%
The framework also facilitates comparison of two or more cyber-physical systems against a given threat model.

\section*{Acknowldegments}
%The work was supported by the 
%Support of this work by 
The U.S. Department of Energy (DOE) under award DE-OE0000895 and the National Science Foundation (NSF) support under awards ECCS-1916142 and ECCS-1808064, and the work of the CYPRES team, are gratefully acknowledged.
%, and the Sandia National Laboratories’ directed R\&D projects.
%\#222444 and \#222444.  
%The data that support the findings of this study are available in \url{https://katedavis.engr.tamu.edu/projects/defenda/

\vspace{-1mm}

% funds from the US Department of Energy under award DE-OE0000895
%\footnotesize{}

%\section{References} 

%\IEEEtriggeratref{12}
\bibliographystyle{IEEEtran}
\bibliography{kreferences}
%\section*{}

\small

\end{document}